\documentclass[twocolumn,groupedaddress]{revtex4-1}
\usepackage{color}
\usepackage{graphicx}
\usepackage{hyperref}
\usepackage{amsmath,amssymb}
\usepackage{epstopdf}
\usepackage{subcaption}
\graphicspath{{figs/}}
\usepackage{dcolumn}

\usepackage{bm}        
\usepackage{verbatim}   
\begin{document}
	
	\title{Quantum Otto and Carnot Cycles via Skew Ising Model}
    \author {Neda Valizadeh}
	\email{valizadeh.neda@email.kntu.ac.ir}
	\affiliation{Department of Physics, University of Mohaghegh Ardabili, P.O. Box 179, Ardabil, Iran}
	\author {Nayyere Einali Saghavaz}
	\affiliation{Department of Physics, University of Mohaghegh Ardabili, P.O. Box 179, Ardabil, Iran}
	\author {Zahra Ebadi}
	\affiliation{Department of Physics, University of Mohaghegh Ardabili, P.O. Box 179, Ardabil, Iran}
	\author {Hosein Mohammadzadeh}
	\affiliation{Department of Physics, University of Mohaghegh Ardabili, P.O. Box 179, Ardabil, Iran}
	

	\begin{abstract}
		We investigate the thermodynamic performance of quantum heat engines and refrigerators based on a two-spin system subject to a skew magnetic field. The working substance is described by an interacting spin model that incorporates both spin--spin coupling and anisotropy induced by a tilted magnetic field. We analyze and compare quantum Carnot and Otto cycles, showing that the Carnot cycle exhibits a universal, entropy-driven behavior with smooth phase boundaries, while the Otto cycle displays a much richer structure governed by the interplay between the energy spectrum and nonequilibrium population differences. In particular, we identify a crossover in both efficiency and coefficient of performance as a function of the interaction strength, which arises from the competition between the interaction energy scale and the magnetic field. We further demonstrate that the skew angle induces state hybridization, modifying both the energy levels and occupation probabilities. Our results highlight that interactions and anisotropy, when properly tuned, can enhance thermodynamic performance, and emphasize the importance of multi-level effects in the design of quantum thermal machines.
	\end{abstract}

	\maketitle

	\section{Introduction}\label{}

The rapid development of quantum technologies has stimulated growing interest in quantum thermodynamics\cite{deffner2019quantum,watanabe2017quantum,vinjanampathy2016quantum,ma2017quantum}, particularly in the design and analysis of quantum heat engines and refrigerators\cite{dann2020quantum,del2022quantum,myers2022quantum,camati2019coherence,chen2020power}. Unlike their classical counterparts, quantum thermal machines operate with working substances governed by discrete energy spectra, quantum coherence, and correlations, such as two-level systems \cite{quan2009quantum,quan2007quantum,huang2012effects,huang2014special,chand2017measurement,thomas2011coupled}, multilevel systems \cite{quan2005quantum,bender2000quantum,bender2002entropy}, harmonic oscillators \cite{quan2009quantum,quan2007quantum,thomas2017implications,wang2015efficiency,rezek2006irreversible}, single ions \cite{abah2012single,rossnagel2014nanoscale}, two ions \cite{chand2017measurement}, quantum dots \cite{sothmann2012magnon}, opto-mechanical systems \cite{yoder2005opto}, and cold bosons \cite{fialko2012isolated}. These intrinsically quantum features open new possibilities for controlling thermodynamic performance, while simultaneously introducing complexities that go beyond classical descriptions~\cite{Vinjanampathy2016, Goold2016}.\\
A central goal in this field is to understand how microscopic properties of the working substance influence macroscopic thermodynamic quantities such as work, heat, efficiency, and coefficient of performance. In this context, quantum heat engines provide a natural framework for investigating the interplay between quantum mechanics and thermodynamics, as they employ quantum systems to convert heat into useful work~\cite{abd2023comparative,quan2007quantum,beretta2012quantum}. In particular, spin-based models have attracted considerable attention due to their conceptual simplicity, controllability, and potential experimental realizations. These systems allow independent tuning of interactions, external fields, and anisotropy, making them ideal platforms for exploring quantum thermodynamic behavior.\\
In recent years, the role of interaction in quantum Carnot and Otto cycles has been extensively explored using coupled spin working substances. It has been shown that even minimal models such as two interacting spins can exhibit rich thermodynamic behavior, including efficiency enhancement, performance optimization, and interaction-driven phase-like transitions in work output~\cite{valizadeh2025two, purkait2023measurement,piccitto2022ising,xi2017quantum,ahadpour2021coupled,huang2013special,he2012thermal,he2012entangled,asadian2022quantum}. These studies demonstrate that the Ising coupling plays a crucial role in determining the energy level structure and hence directly affects heat and work exchange processes during thermodynamic cycles.\\
Among the various thermodynamic cycles, the Carnot and Otto cycles play a fundamental role. The quantum Carnot cycle consists of two isothermal and two adiabatic processes and represents an ideal reversible engine operating between two heat reservoirs. Its efficiency is given by the Carnot bound
\begin{equation}
\eta_C = 1 - \frac{T_{\mathrm{cold}}}{T_{\mathrm{hot}}},
\end{equation}
which depends only on the reservoir temperatures and is independent of the microscopic details of the working substance~\cite{quan2007quantum,beretta2012quantum}. However, its practical realization is highly challenging due to the requirement of quasi-static control.\\
In contrast, the quantum Otto cycle consists of two isochoric and two adiabatic processes and is more feasible for experimental implementation. In this cycle, heat exchange occurs during the isochoric strokes, while work is performed during the adiabatic transformations. Unlike the Carnot cycle, the performance of the Otto cycle depends explicitly on the energy spectrum and nonequilibrium population distributions, making it highly sensitive to system parameters and a powerful probe of microscopic effects~\cite{RezekKosloff2006,quan2007quantum,beretta2012quantum}.\\
Recent studies have demonstrated that interactions and anisotropy can significantly modify the thermodynamic performance of quantum heat engines. Spin--spin coupling can induce nontrivial spectral restructuring, while transverse fields lead to hybridization of eigenstates and redistribution of occupation probabilities. Despite these advances, a comprehensive understanding of how these competing mechanisms jointly determine the operation of quantum thermodynamic cycles, particularly in multi-level systems, remains incomplete.

In this work, we address this problem by considering a two-spin system subject to a skew magnetic field as the working substance of quantum Carnot and Otto cycles. This model incorporates both interaction and anisotropy in a controllable manner, allowing us to investigate the interplay between the coupling strength $J$, the magnetic field $h$, and the skew angle $\alpha$. We analyze the work output, efficiency, and coefficient of performance under different operating conditions, and compare the results with those of a non-interacting two-level system.\\
Our results show that while the Carnot cycle exhibits a smooth and universal phase structure governed by entropy, the Otto cycle reveals a much richer behavior arising from the competition between energy level structure and population redistribution. In particular, we identify a crossover in thermodynamic performance as a function of the interaction strength, which can be understood in terms of the competition between interaction and magnetic energy scales.\\
The paper is organized as follows. In Sec.~\ref{sec:SPIN-WORKING}, we present the model Hamiltonian and discuss the spectral and thermal properties of the working substance. The formulation of the quantum Carnot and Otto cycles is introduced in Secs.~\ref{carno cycle} and \ref{otto cycle}. The corresponding thermodynamic behavior and performance are analyzed in Sec.~\ref{results}, where the results and discussion are presented in detail. Finally, the main findings are summarized in the Conclusions section \ref{sec:conclusions}.

\section{Spin Working Substance and Considered Cycles}
\label{sec:SPIN-WORKING}

We consider a two-spin system as the working substance of the thermodynamic cycles. The system is described by an interacting spin model subject to a skew magnetic field. The Hamiltonian reads
\begin{equation}
H = -J S_z^{(1)} S_z^{(2)} - h \sin(\alpha)\, S_x - h \cos(\alpha)\, S_z,
\label{eq:hamiltonian}
\end{equation}
where $J$ is the spin--spin coupling constant and $S_x$, $S_z$ denote collective spin operators. The external magnetic field has a fixed magnitude $h = |h|$ and is oriented at an angle $\alpha \in [0,\pi/2]$ with respect to the $z$-axis:
\begin{equation}
\mathbf{h} = h \sin(\alpha)\, \hat{i} + h \cos(\alpha)\, \hat{k}.
\end{equation}

The sign of $J$ determines the nature of the interaction: $J>0$ corresponds to antiferromagnetic coupling, while $J<0$ describes ferromagnetic interaction. The skew angle $\alpha$ introduces anisotropy through the transverse field component, which leads to hybridization of spin states.\\

The thermodynamic properties are derived from the partition function
\begin{equation}
Z = \sum_{i=1}^{4} e^{-\beta E_i},
\label{eq:partition}
\end{equation}
where $\beta = 1/(k_B T)$.

We consider spin-$\frac{1}{2}$ particles, for which the Hilbert space is four-dimensional. The energy spectrum $\{E_i\}$ is obtained by diagonalizing the Hamiltonian in Eq.~(\ref{eq:hamiltonian}). The eigenvalues can be expressed as

\begin{align}
E_1 &= J, \\
E_2 &= \frac{a - J}{3} - \frac{4\left(-h^2 - \frac{J^2}{3}\right)}{a}, \\
E_3 &= \frac{1}{2} \left( \sqrt{4h^2 + J^4 - J E_2 + (J+E_2)^2 - E_2^2 - 4} - J - E_2 \right), \\
E_4 &= -J - E_2 - E_3,
\end{align}

The parameter $a$ entering the eigenvalues is given by
\begin{widetext}
\begin{equation}
a = \left( 36 h_x^2 J - 72 h_z^2 J + 8 J^3 
+ 12 \sqrt{
-12 h_x^6 - 36 h_x^4 h_z^2 - 36 h_x^2 h_z^4 - 12 h_z^6 
- 3 h_x^4 J^2 - 60 h_x^2 h_z^2 J^2 
+ 24 h_z^4 J^2 - 12 h_z^2 J^4
} \right)^{1/3},
\end{equation}
\end{widetext}

where, for compactness, we have used $h_x = h \sin\alpha$ and $h_z = h \cos\alpha$.

The corresponding eigenstates $\{|E_i\rangle\}$ form a complete orthonormal basis of the Hilbert space and are obtained from the diagonalization of the Hamiltonian. Due to the transverse field component, these eigenstates are, in general, superpositions of the computational basis states.

At thermal equilibrium, the occupation probability of each level is given by the Boltzmann distribution
\begin{equation}
P_i = \frac{e^{-\beta E_i}}{Z}, \quad i=1,\dots,4,
\label{eq:prob}
\end{equation}

In the following, we set $k_B = 1$, so that $\beta = 1/T$. The set of probabilities $\{P_i\}$ fully characterizes the thermal state of the system and determines the heat and work exchanged during the thermodynamic cycles.

\subsection{Modeling of quantum Carnot cycle}\label{carno cycle}

The modeling of the quantum Carnot cycle is a theoretical framework that aims to understand and analyze the behavior of a Carnot cycle operating in the quantum regime. The Carnot cycle is a thermodynamic cycle that represents the most efficient way to convert heat into work, and it consists of four processes: isothermal expansion, adiabatic expansion, isothermal compression, and adiabatic compression \cite{quan2009quantum}.
Using the heat exchange $dQ = TdS$ \cite{quan2014maximum} in the two quantum processes with constant temperature, the heat that is absorbed from the hot reservoir $Q_{1}$ and the heat that is given to the cold reservoir $Q_{2}$ are calculated, respectively, by
\begin{eqnarray}
Q_{1}&=&T_{1}(S_{2}-S_{1}) \label{Q1}\\
Q_{2}&=&T_{2}(S_{4}-S_{3})
\label{Q2}
\end{eqnarray}

where $T_{1}$ and $T_{2}$ are the temperatures of the hot and cold reservoirs respectively. The entropy is given by
		\begin{equation}
			S_{i}=-\sum_{n=1}^{4}P_{n}\ln P_{n}.
		\end{equation}
Where $P_{n}$ due to the  \ref{eqt:probabilities} is the Boltzmann distribution of a thermal equilibrium state
and the first law of thermodynamics we obtain the net work done during a quantum Carnot cycle 
\begin{equation}
W=Q_{1}+Q_{2},
\end{equation}
Next, we examine the operation efficiency ($\eta$) of the quantum Carnot cycle,
\begin{equation}
\eta=\frac{W}{Q_{1}}=1-\frac{T_{2}}{T_{1}}.
\end{equation}

Now we analyze for Skew model:\\
The entropies of the working substance at different instants $i=1,2,3,4$ are evaluated using the occupation probabilities . The external magnetic field at isothermal process $T_{1}$ is $h_{1}$, occupation probabilities are $P_{1},P_{2},P_{3},P_{4}$ while at isothermal process $T_{2}$ is $h_{2}$ and occupation probabilities are $ P_{1}^{'}, P_{2}^{'}, P_{3}^{'}, P_{4}^{'}$.
\begin{eqnarray}
S_{1}=-P_{1}\ln P_{1}-P_{2}\ln P_{2}-P_{3}\ln P_{3}-P_{4}\ln P_{4}.\\
S_{2}=-P_{1}^{'}\ln P_{1}^{'}-P_{2}^{'}\ln P_{2}^{'}-P_{3}^{'}\ln P_{3}^{'}-P_{4}^{'}\ln P_{4}^{'}.
\end{eqnarray}

The net work for this model done during quantum Carnot cycle is
\begin{equation}
W=Q_{1}+Q_{2}=(T_{1}-T_{2})(S_{2}-S_{1}),
\label{Eq:W}
\end{equation}
where we have used the relations $ S_{2}= S_{3}$ and $ S_{1}= S_{4}$. This equality is because of the fact that the occupation probabilities and thus the entropy is constant in any quantum adiabatic process.

\begin{figure*}[t]
\begin{center}
\includegraphics[width=16cm, clip]{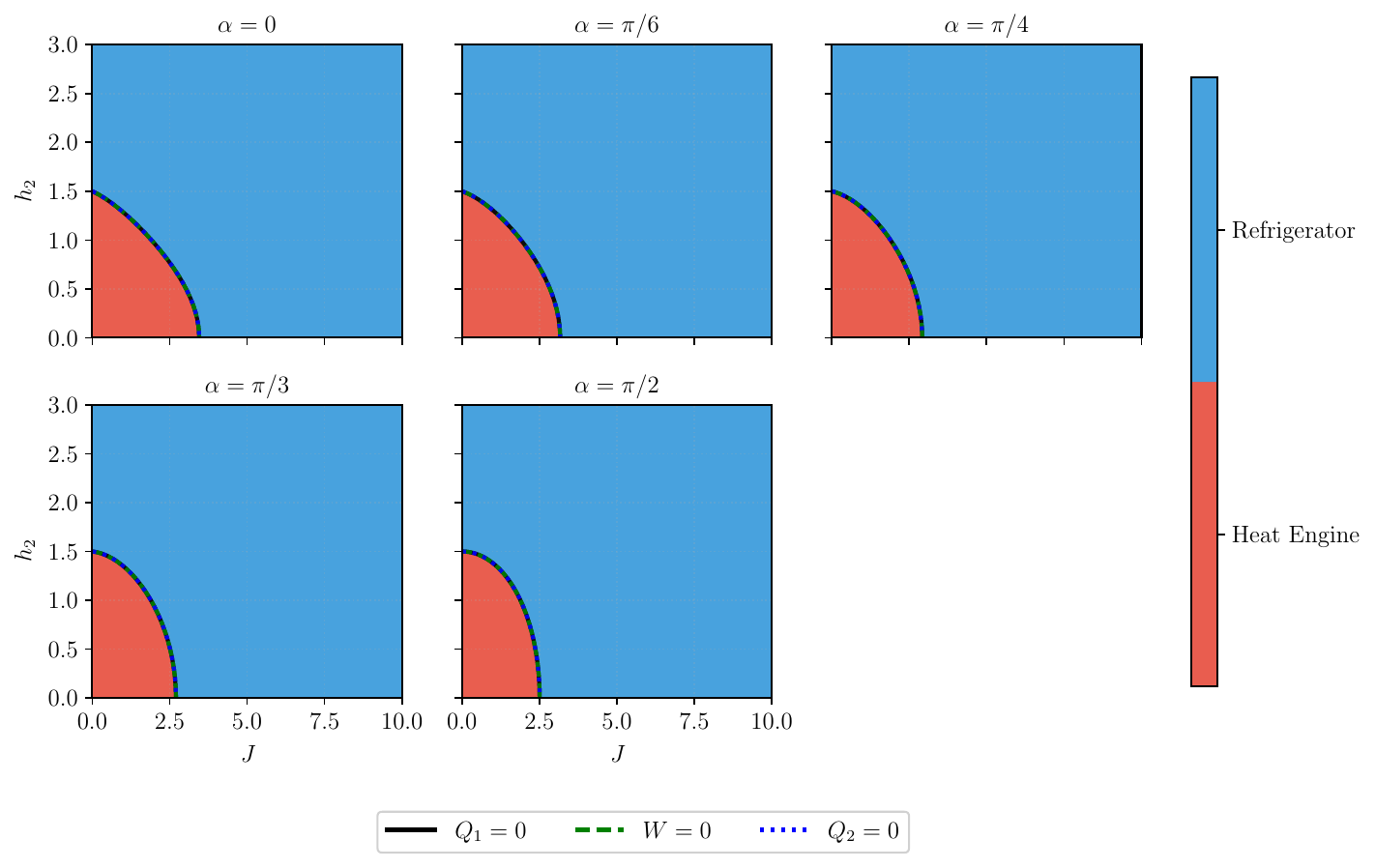}
\caption{Operational phase diagram of the quantum Carnot cycle using the two-spin Skew model as the working substance. The diagram shows the heat engine and refrigerator regimes as functions of the external magnetic field \(h_2\) and the coupling constant \(J\) for fixed parameters \(h_1 = 3\), \(T_1 = 10\), \(T_2 = 5\), and skew angle \(\alpha \in [0,\pi/2]\). The phase boundaries are determined by the condition \(\Delta S = 0\), and the distinction between engine and refrigerator regions is indicated directly in the figure.}
\label{Carno}
\end{center}
\end{figure*}

\begin{figure*}[t]
\centering
\includegraphics[width=15cm, clip]{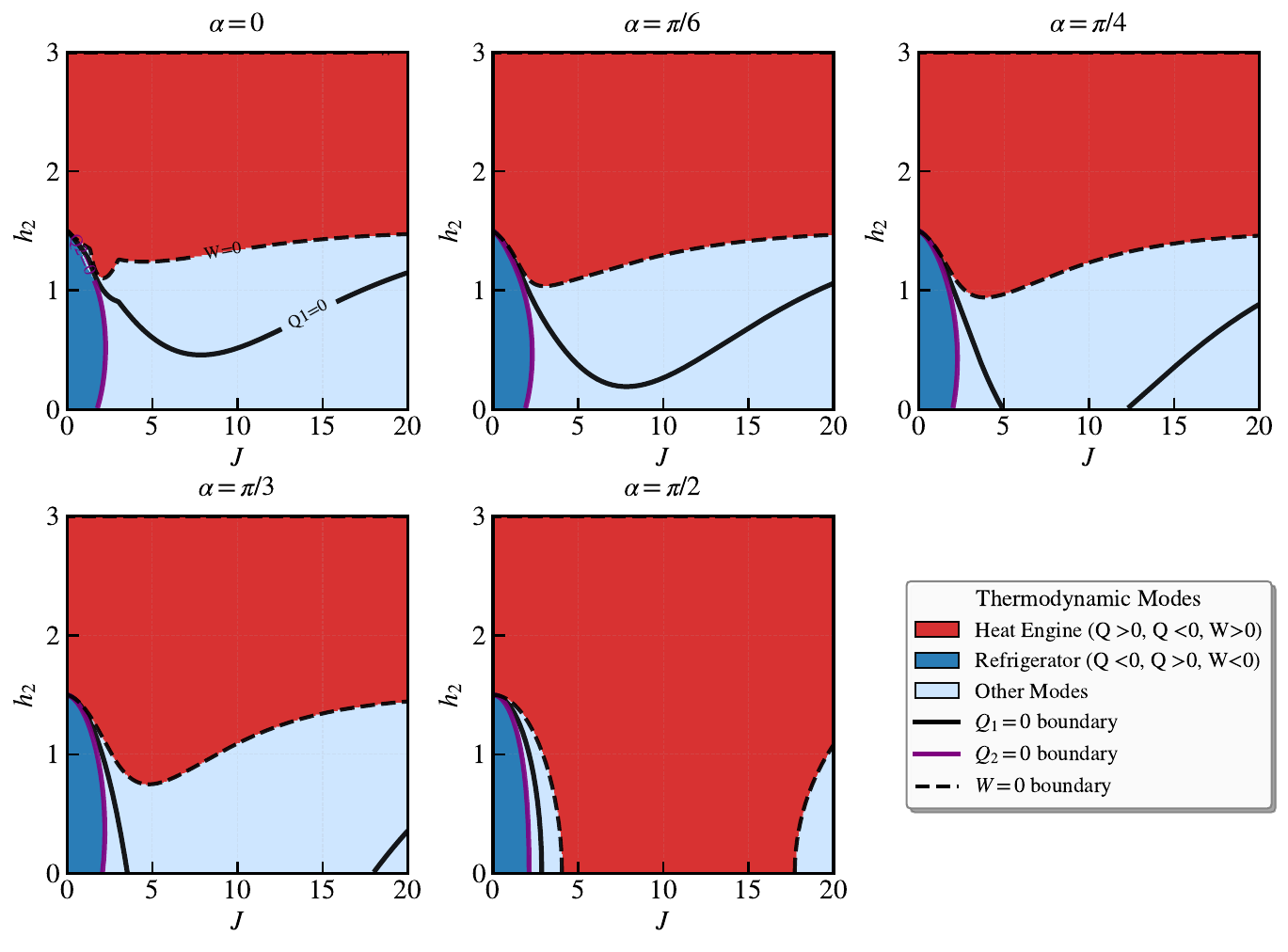}
\caption{Phase diagram of the quantum Otto cycle for the Skew model, showing the heat engine and refrigerator regimes for various skew angles \(\alpha\). Each panel corresponds to a different value of \(\alpha\) (from \(\alpha = 0\) to \(\alpha = \pi/2\)), with the engine regime defined by \(Q_1 > 0\), \(Q_2 < 0\), \(W > 0\) and the refrigerator regime by \(Q_1 < 0\), \(Q_2 > 0\), \(W < 0\). Other parameters are \(h_1 = 3\), \(T_1 = 10\), and \(T_2 = 5\). The distinction between regimes is indicated directly within each panel.}
\label{otto}
\end{figure*}

\subsection{Modeling of quantum Otto cycle}\label{otto cycle}

The study of the quantum Otto cycle is a fundamental topic in quantum thermodynamics, exploring the operational principles of heat engines and refrigerators that utilize quantum systems as working substances. The cycle consists of four distinct stages: two isochoric processes, where heat is exchanged with thermal reservoirs at different temperatures, and two quantum adiabatic processes, where the system's internal parameters (such as the external magnetic field or coupling constants) are modulated without heat exchange. This framework allows for investigating the influence of quantum coherence, entanglement, and strong coupling on the performance metrics of quantum heat engines \cite{kosloff2017quantum, das2020quantum, leggio2016otto}.

During the isochoric stages, the heat absorbed from the hot reservoir ($Q_1$) and the heat rejected to the cold reservoir ($Q_2$) are determined by the change in the occupation probabilities of the energy levels:
\begin{align}
Q_{1} &= \sum_{n=1}^{4} E_{n} (P_{n} - P_{n}'), \label{Q1O} \\
Q_{2} &= -\sum_{n=1}^{4} E_{n}' (P_{n} - P_{n}'), \label{Q2O}
\end{align}
where $E_n$ ($E_n'$) denotes the $n$-th eigenenergy of the Hamiltonian during the hot (cold) isochoric process. According to the first law of thermodynamics, the net work $W$ performed by the Otto engine over a complete cycle is given by:
\begin{equation}
W = Q_{1} + Q_{2}.
\label{WO}
\end{equation}

For the specific case of the "Skew model," the heat exchange expressions can be explicitly written in terms of the four-level populations as:
\begin{align}
Q_{1} &= E_{1}(P_{1}-P_{1}') + E_{2}(P_{2}-P_{2}') + E_{3}(P_{3}-P_{3}') + E_{4}(P_{4}-P_{4}'), \\
Q_{2} &= -E_{1}'(P_{1}-P_{1}') - E_{2}'(P_{2}-P_{2}') - E_{3}'(P_{3}-P_{3}') - E_{4}'(P_{4}-P_{4}').
\end{align}
Consequently, the total work done during the cycle simplifies to:
\begin{equation}
W = 2(h_{2}-h_{1}) \left[ (P_{3}-P_{3}') + (P_{4}' - P_{4}) \right].
\end{equation}

The thermal efficiency of the engine is defined as $\eta = \frac{W}{Q_{1}}$. In the uncoupled limit (where inter-spin interactions vanish), the efficiency reduces to the standard form:
\begin{equation}
\eta_{0} = 1 - \frac{h_{2}}{h_{1}}.
\end{equation}

\begin{figure*}
			\begin{subfigure}{0.32\textwidth}\includegraphics[width=\textwidth]{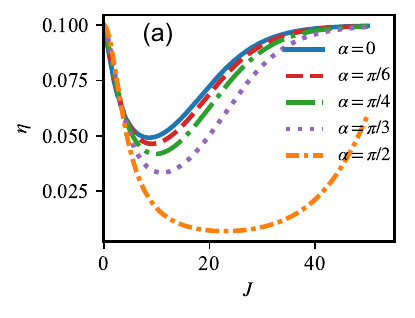}
                \label{1}
			\end{subfigure}
			\begin{subfigure}{0.32\textwidth}\includegraphics[width=\textwidth]{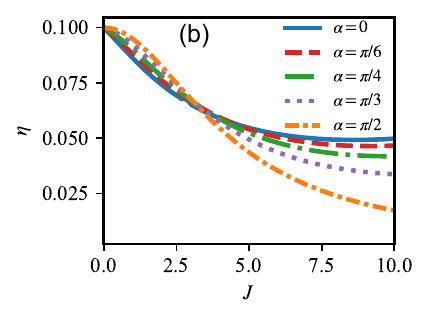}
                    \label{2}
			\end{subfigure}
			\begin{subfigure}{0.32\textwidth}\includegraphics[width=\textwidth]{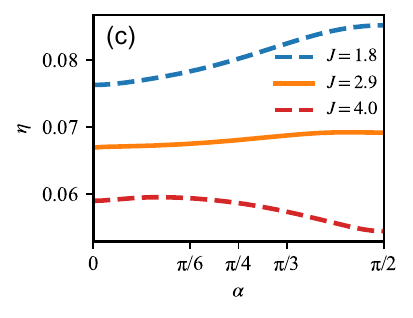}
                   \label{3}
			\end{subfigure}
			
			\caption{Efficiency \(\eta\) of the quantum Otto cycle as a function of the coupling constant \(J\). The efficiency exhibits a non-monotonic behavior with a minimum at an intermediate value \(J^*\), indicating a crossover between weak- and strong-coupling regimes. Parameters: \(h_1=3\), \(T_1=10\), \(T_2=5\), and several values of \(\alpha\).}
			\label{eff}
		\end{figure*}

\begin{figure*}
			\begin{subfigure}{0.45\textwidth}\includegraphics[width=\textwidth]{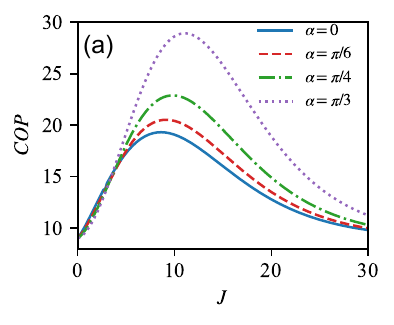}
                \label{1}
			\end{subfigure}
			\begin{subfigure}{0.45\textwidth}\includegraphics[width=\textwidth]{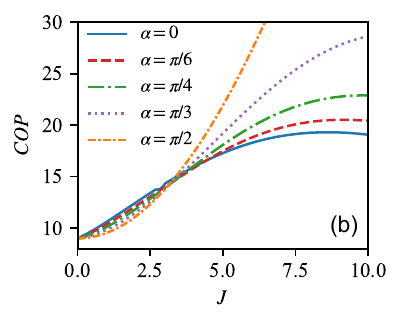}
                    \label{2}
			\end{subfigure}
			\caption{Coefficient of performance (COP) of the quantum Otto cycle as a function of the coupling constant \(J\). The COP shows complementary behavior to the efficiency, with maxima corresponding to optimal refrigeration performance. The interplay between the interaction scale \(J\) and the magnetic scale \(h\) determines the location of the maxima. Parameters are the same as in Fig.~3.}
			
			\label{cop}
		\end{figure*}

\section{Results and Discussion}\label{results}

The results presented in Figs.~\ref{Carno}--\ref{otto} reveal how interaction, anisotropy, and thermal populations jointly determine the thermodynamic performance of the Skew model. The ordering of figures highlights the transition from equilibrium-controlled behavior (Carnot cycle) to nonequilibrium and spectrum-dependent features (Otto cycle), followed by performance characteristics.

Before proceeding to the analysis of the results, it is important to clarify the role of the external magnetic field $h$ in the present model. While the coupling constant $J$ controls the interaction strength and the skew angle $\alpha$ introduces anisotropy and state hybridization, the magnetic field sets the fundamental energy scale of the system.

All eigenvalues $E_n = E_n(J,h,\alpha)$ depend explicitly on $h$, and therefore the level spacing and thermal occupation probabilities are directly governed by its magnitude. In the Otto cycle, the field also plays the role of the driving parameter, since the work originates from its modulation between $h_1$ and $h_2$. Consequently, the thermodynamic behavior of the system is determined by the interplay between the interaction scale $J$ and the magnetic scale $h$, with $\alpha$ controlling the degree of mixing between the eigenstates.

\subsection{Carnot Phase Diagram}

We begin with the phase diagram of the quantum Carnot cycle shown in Fig.~\ref{Carno}. The boundaries separating the heat engine and refrigerator regimes are smooth and relatively regular, reflecting the equilibrium nature of the Carnot cycle.

Since the net work is given by Eq(\ref{Eq:W}),
the operational mode is entirely determined by the entropy difference $\Delta S$. Therefore, the phase boundaries correspond to $\Delta S = 0$ independent of dynamical details.

From a microscopic perspective, the entropy depends on the Boltzmann distribution over the energy levels. The parameters $J$ and $\alpha$ affect the entropy indirectly by modifying the eigenvalue spectrum and lifting degeneracies. However, because the Carnot cycle depends only on equilibrium properties, the phase diagram remains relatively insensitive to fine spectral rearrangements.

The skew angle $\alpha$ introduces state mixing through the transverse field component, leading to smooth deformations of the entropy landscape. Similarly, $J$ modifies the level spacing and correlations, but its effect is reflected only through equilibrium population redistribution. This explains the regular geometry of the phase boundaries.

\subsection{Otto Phase Diagram}

In contrast, the phase diagram of the Otto cycle shown in Fig.~\ref{otto} exhibits a much richer and more intricate structure. This complexity arises because the Otto cycle depends explicitly on both the energy spectrum and the nonequilibrium population differences.
The expressions for heat and work are provided in equations (\ref{Q1O}), (\ref{Q2O}), and (\ref{WO}), the operational regime is determined by weighted sums over all energy levels.

The phase boundaries correspond to nontrivial conditions such as
\begin{equation}
Q_1 = 0, \quad Q_2 = 0, \quad W = 0,
\end{equation}
each involving a delicate balance between energy differences and population imbalances.

As a result, small changes in $J$ or $\alpha$ can significantly alter the relative contributions of different levels, leading to strongly deformed and highly structured phase boundaries.

The coupling constant $J$ reshapes the spectrum and can induce substantial changes in level spacing, while the skew angle $\alpha$ enhances hybridization and redistributes population weights. Unlike the Carnot case, the phase diagram here reflects a genuine competition between spectral structure and nonequilibrium population dynamics.

The efficiency of the Otto cycle, shown in Fig.\ref{eff}, exhibits a clear non-monotonic dependence on the coupling constant $J$, with a minimum at an intermediate value $J^*$. This behavior signals a crossover between two distinct regimes.

In the weak-coupling regime, the interaction acts perturbatively, leading to level mixing and reduced energy selectivity, which diminishes the efficiency. In the strong-coupling regime, the spectrum is reorganized into interaction-dominated branches, enhancing the relevant energy differences and improving work extraction.

Importantly, this crossover does not correspond to a simple dominance shift between low- and high-energy levels. Instead, it reflects a redistribution of contributions across multiple levels, governed by the interplay between energy differences and population imbalances.

The skew angle $\alpha$ smoothens the crossover by increasing state hybridization, thereby reducing the contrast between weak- and strong-coupling regimes.

The coefficient of performance (COP), shown in Fig.~\ref{cop}, exhibits complementary behavior, with maxima corresponding to optimal refrigeration performance.

From a microscopic viewpoint, the COP is maximized when the population differences $\Delta P_n$ are optimally aligned with the energy spectrum, allowing efficient heat extraction from the cold reservoir with minimal work input. As in the efficiency case, the COP is determined by a balance between spectral properties and population redistribution.

The dependence on $J$ and $\alpha$ confirms that neither the energy levels nor the occupation probabilities alone determine the performance. Instead, the thermodynamic behavior emerges from their combined and parameter-dependent contributions.


\section{Conclusions}\label{sec:conclusions}

In this work, we have investigated the thermodynamic performance of quantum Carnot and Otto cycles utilizing a two-spin Skew model as the working substance. By incorporating spin--spin interactions and a skewed magnetic field, this model captures the complex interplay between correlations, anisotropy, and multi-level spectral structures in quantum thermal machines.

Our results demonstrate a fundamental qualitative distinction between the two cycles. The Carnot cycle is governed by equilibrium thermodynamics, with its operational boundaries determined solely by entropy variations. Consequently, its performance remains smooth and relatively insensitive to fine-scale spectral rearrangements, consistent with the universal character of Carnot-limited engines.

In contrast, the Otto cycle exhibits a pronounced sensitivity to both the energy spectrum and nonequilibrium population dynamics. The resulting phase diagrams are significantly more intricate, emerging from a delicate balance between energy-level shifts and population redistribution. This underscores the microscopic nature of the Otto cycle, where the specific structure of the Hamiltonian directly dictates the performance metrics.

A key finding of this study is the identification of a performance crossover in both efficiency and the coefficient of performance (COP) as a function of interaction strength. This crossover signifies a competition between the interaction energy scale $J$ and the magnetic energy scale $h$. In the weak-coupling regime, interaction-induced state mixing tends to diminish thermodynamic performance. Conversely, in the strong-coupling regime, the interaction-driven spectral restructuring enhances effective energy gaps, thereby improving work extraction and cooling capacity.

We have further demonstrated that the skew angle $\alpha$ plays a pivotal role by inducing state hybridization through the transverse field component. This effect smoothens the crossover and modulates the operational regimes by redistributing population weights among the eigenstates.

In conclusion, our findings suggest that interactions and anisotropy, when properly tuned, can be leveraged to enhance rather than degrade the performance of quantum thermal machines. This provides a clear physical guideline for the design of quantum devices: optimal operation is achieved by exploiting the competition between interaction and magnetic energy scales, rather than simply minimizing interactions.

	\bibliography{refs}
	\end{document}